\documentclass[%
aps,
pra,
amsmath,amssymb,
reprint,
]{revtex4-2}

\usepackage{graphicx}
\usepackage{dcolumn}
\usepackage{bm}
\usepackage[utf8]{inputenc}
\usepackage[T1]{fontenc}
\usepackage{etoolbox}
\usepackage{amsmath}
\usepackage{amssymb}
\usepackage{graphicx}
\usepackage[colorlinks=true, allcolors=blue]{hyperref}
\usepackage{gensymb}
\usepackage{xcolor}

\begin{document}

\title{Airborne Minnaert-Like Resonance of an Air-Filled Elasto-Bubble}

\author{F. Delmotte}%
\affiliation{Universit\'e Paris Cit\'e, CNRS, Laboratoire Mati\`ere et Syst\`emes Complexes UMR 7057, 75231 Paris cedex 13, France}%
\author{V. Leroy}
\affiliation{Universit\'e Paris Cit\'e, CNRS, Laboratoire Mati\`ere et Syst\`emes Complexes UMR 7057, 75231 Paris cedex 13, France}%
\author{J. Zhang}
\email{jishen.zhang@espci.fr}
\affiliation{Universit\'e Paris Cit\'e, CNRS, Laboratoire Mati\`ere et Syst\`emes Complexes UMR 7057, 75231 Paris cedex 13, France}%
\affiliation{PMMH, CNRS, ESPCI Paris, Universit\'e PSL, Sorbonne Universit\'e, Universit\'e Paris Cit\'e, F-75005, Paris, France.}%

\date{\today}

\begin{abstract}
Deep-subwavelength acoustic resonators are key building blocks of acoustic metamaterials, yet achieving bubble-like resonances in air remains challenging because the Minnaert mechanism relies on the inertia of a surrounding liquid. Here we demonstrate that air-filled soft elastomer shells, termed elasto-bubbles, realize an airborne analogue of the Minnaert resonator. Using impedance-tube measurements together with the theory of layered-bubble scattering, we show that these soft hollow capsules sustain strong monopolar resonances despite being deeply subwavelength. Their resonance frequency, transmission dip, and absorption are quantitatively captured, without adjustable parameters, by a model accounting for shell elasticity and viscoelasticity. Because shell radius and thickness can be tuned independently during fabrication, elasto-bubbles provide a simple and versatile platform for airborne acoustic metamaterials, resonant absorbers, and acoustic filters.
\end{abstract}

\maketitle
Metamaterials are artificial structures that exhibit unusual wave properties, such as negative refraction~\cite{smith2004metamaterials}. They are generally built from subwavelength resonators, which must simultaneously interact strongly with the incident wave while remaining sufficiently small to enable an effective-medium description. For electromagnetic waves, the paradigmatic example is the split-ring resonator~\cite{pendry1999magnetism}. In acoustics~\cite{ma2016acoustic}, representative resonant building blocks include Helmholtz resonators~\cite{fang2006ultrasonic}, membranes~\cite{lee2010composite}, and bubbles~\cite{leroy2015superabsorption}.

Among these, gas bubbles are particularly remarkable because their Minnaert resonance occurs at wavelengths orders of magnitude larger than their radius~\cite{minnaert1933}. A useful quantity to characterize this strong subwavelength character is the product of the resonance frequency and resonator radius, $f_0R$, which can be compared with the sound speed in the surrounding medium. For a bubble in water resonating at the Minnaert frequency, one has $f_0R=3.2\,$m/s~\cite{minnaert1933}, to be compared with a sound speed of about $1500\,$m/s in water, corresponding to a wavelength-to-radius ratio close to 500. This exceptional scale separation has enabled bubble-based metamaterial concepts for water-borne acoustics in both the ultrasonic and audible frequency ranges~\cite{leroy2015superabsorption}.

Transposing this physical mechanism to airborne acoustics is, however, not straightforward. Bubble resonance can be interpreted as a mass--spring oscillator in which gas compressibility provides the restoring force while the surrounding medium provides inertia. In water, the displaced liquid supplies a large inertial load, whereas in air the surrounding medium is too light to play this role efficiently. This raises the question of whether an airborne analogue of bubble resonance can be realized by replacing the inertia of the surrounding liquid with that of a thin dense shell.

In this work, we investigate such a system using an air-filled elastomeric shell, which we term an elasto-bubble. In this object, the compressible gas core provides the restoring force while the shell supplies inertia, thereby reproducing the mass--spring mechanism underlying Minnaert resonance. Achieving this regime requires a shell that is sufficiently dense to provide inertial loading while remaining compliant enough that its own elastic stiffness does not dominate the gas compressibility. Soft elastomers satisfy these two conditions simultaneously, making them well suited for this purpose. Beyond its fundamental interest as an airborne analogue of the bubble resonator, such a structure could provide a simple route toward compact resonant inclusions for airborne acoustic metamaterials, filters, and absorbers.

The acoustic response of this object can be described using the theory developed by Alekseev and Rybak for a gas bubble surrounded by an elastic shell in an elastic medium~\cite{alekseev1999gas}. Let us consider a gas bubble of radius $R$, surrounded by a shell of thickness $h$ made of a material with density $\rho$ and shear modulus $\mu=\mu^\prime +\mathrm{i}\mu^{\prime\prime}$. The embedding medium is air, with density $\rho_a$ and wavenumber $k_a$. In the long-wavelength limit, $k_aR\ll1$, an incident plane wave $p\exp(\mathrm{i}k_ax)$ of amplitude $p$ gives rise to a scattered spherical wave $f_\text{scat}p\exp(\mathrm{i}k_ar)/r$, where the scattering function reads
\begin{eqnarray}
f_\text{scat}=\frac{R^\star}{(\omega_0/\omega)^2-1-\mathrm{i} (\delta +k_aR^\star)}, \label{eq0}
\end{eqnarray}
with, at first order in $h/R$,
\begin{eqnarray}
R^\star&=&R\frac{\rho_aR+\rho_a h}{\rho_a R+\rho h},\label{eq1} \\
\omega_0^2 &=& \frac{3\kappa P_0 +12\mu^\prime h/R}{\rho_a R^3}R^\star,\label{eq2}\\
\delta &=& \frac{12\mu^{\prime\prime} h/R}{\omega^2\rho_a R^3}R^\star, \label{eq3}
\end{eqnarray}
where $\kappa$ is the polytropic exponent for the gas. Here, $R^\star$ is an effective acoustic radius that accounts for the inertia of the shell, $\omega_0$ is the resonance angular frequency of the elasto-bubble, and $\delta$ is a dimensionless damping coefficient associated with viscoelastic dissipation in the shell. If the density contrast between the shell and the gas is higher than the radius-to-thickness ratio of the elasto-bubble ($\rho h/(\rho_a R)\gg 1$), the effective radius can be approximated by $R^\star\simeq \rho_a R^2/(\rho h)$. If furthermore the shear modulus of the shell is low enough to ensure that $\mu^\prime h/R\ll \kappa P_0/4$, the resonance frequency reduces to
\begin{eqnarray}
\omega_0^2 &=& \frac{3\kappa P_0}{\rho Rh}.\label{eq2b}
\end{eqnarray}

This simplified expression highlights the close analogy with Minnaert resonance while making explicit the replacement of liquid inertia by shell inertia. It predicts a resonance frequency of the form $f_0R=\sqrt{R/h}\times 3.2\,$m/s. As expected, the resonance is not as low as for a bubble in water, because only a finite shell contributes to the inertia. Nevertheless, the system remains strongly subwavelength. For $R/h=10$, one obtains $f_0R\simeq 10\,$m/s, corresponding to an acoustic wavelength in air about 34 times larger than the elasto-bubble radius. This level of scale separation is already sufficient to make such objects appealing as airborne resonant inclusions.

\begin{figure}
\includegraphics[width=0.9\linewidth]{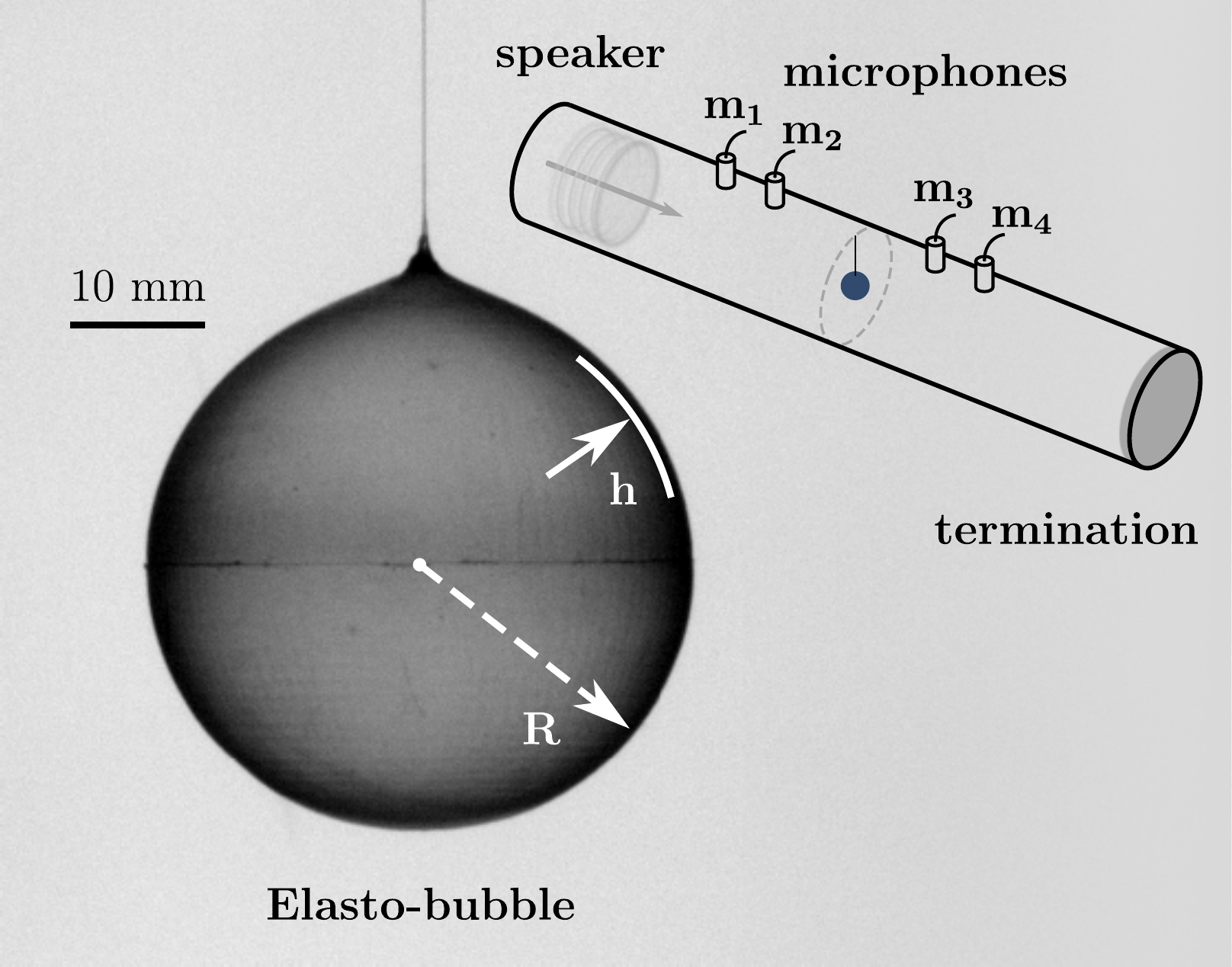}
\caption{\label{fig:schema} Representative elasto-bubble} suspended on a thin wire. The annotations indicate the radius $R$ of the elasto-bubble and the shell thickness $h$. Inset: schematic of the experimental setup.
\end{figure}

To test the validity of Eq.~(\ref{eq2b}), we fabricated elasto-bubbles in the targeted range of radii and shell thicknesses. For a soft elastomer with a density close to that of water, a resonance frequency around $1\,$kHz is expected for $R=20\,$mm and $h=0.5\,$mm. We used an in-house viscous-coating protocol~\cite{eddi2026tunable} that enables independent control of shell radius and thickness. A liquid polymer mixture was first deposited into each hemisphere of a spherical mold and spread over the inner surface. The hemispheres were then placed bottom-down to allow excess liquid to drain. After draining, the mold was sealed and mounted on a planetary rotating platform during curing in order to smooth residual thickness heterogeneities. Once the elastomer was fully cross-linked, the intact shell was demolded while remaining filled with air, without additional inflation, so that in-plane prestress is expected to be negligible. A highly compliant PDMS-based elastomer (Ecoflex 00-30) was used, with an elastic modulus $\mu^\prime\approx 26\,$kPa. The shell thickness was deduced from the elasto-bubble mass through $m=4\pi \rho(R_\mathrm{ext}^3-(R_\mathrm{ext}-h)^3)/3$, with $R_\mathrm{ext}$ the external radius of the elasto-bubble, set by the inner radius of the hemispherical mold, and $\rho=1070\,$kg/m$^3$ the density of the elastomer. Five elasto-bubbles of varying radii and shell thicknesses were manufactured, as listed in Table~\ref{tab:table}, in which we report the theoretical resonance frequency predicted by Eq.\,(\ref{eq2b}). Figure~\ref{fig:schema} shows a representative sample.

\setlength{\tabcolsep}{5pt}
\renewcommand{\arraystretch}{1.2}
\begin{table}[ht]
\centering
\begin{tabular}{|l|c|c||c|c|}
\hline  
No. & $R_\text{ext}$ (mm) & $m$ (g) & $h$ ($\mu$m) & $f_0$ (Hz)\\
 \hline
 1 $\nabla$  & 21     & 1.94      & 332.4 $\pm$ 1.7 & 1211.0 $\pm$ 3.1 \\
  \hline
  2 $\bigcirc$   & 21     & 2.53      & 435.6 $\pm$ 1.8 & 1060.4 $\pm$ 2.1 \\
   \hline
   3 $\square$  & 21     & 3.79      & 659.7  $\pm$ 1.8 & 866.5 $\pm$ 1.1 \\
 \hline
   4 $\triangle$  & 15     & 1.08      & 365.8 $\pm$ 3.5 &1371.8 $\pm$ 6.3 \\
 \hline
   5 $\Diamond$  & 10     & 0.64      & 500.6 $\pm$ 8.2 & 1455.5 $\pm$ 11.3 \\
 \hline
 \end{tabular}
 \caption{\label{tab:table} Parameters of the five elasto-bubbles fabricated and characterized in this study. $R_\mathrm{ext}$ is the mold radius used during fabrication. $m$ is the elasto-bubble mass, from which the shell thickness $h$ is deduced. The theoretical resonance frequency $f_0$ is then evaluated from Eq.~(\ref{eq2b}).}
 \end{table}

The acoustic response of the elasto-bubbles was characterized using a commercial impedance tube (B\&K 4206-T) capable of generating plane acoustic waves in the frequency range $0.3$--$5\,$kHz. The tube has a uniform inner radius $R_\mathrm{t}=50\,$mm (Fig.~\ref{fig:schema}). A speaker is placed at one end of the tube, and four condenser microphones (m1--m4) are positioned along the tube to measure the acoustic field. Two microphones are located upstream of the test section and two downstream. Although the finite tube length leads to multiple reflections, a standard two-termination procedure~\cite{song1999} allows one to determine the transmission and reflection coefficients, $\mathcal{T}$ and $\mathcal{R}$, of the sample placed in the tube center. In our case, the sample is a single elasto-bubble suspended on the tube axis between microphones m2 and m3 using a thin nylon wire of diameter $0.1\,$mm, thereby minimizing the perturbation of the acoustic field.

At rest, the elasto-bubble is free to deform slightly under gravity and adopts a weakly asymmetric pear-like shape (Fig.~\ref{fig:schema}). Because this deformation remains modest, its effect on the monopolar resonance is expected to be secondary compared with uncertainties on shell thickness and viscoelastic parameters. A possible limitation of the experimental protocol is a gradual loss of gas through the PDMS-based shell or through microscopic defects~\cite{goldowsky2013gas, rao2007preparation}. Such a loss would reduce the internal volume and therefore alter the effective radius, potentially affecting the measured acoustic response. To check this point, independent imaging measurements were performed before and after the acoustic experiments. The elasto-bubble shape was extracted from images, allowing the volume to be estimated and an effective radius to be deduced. These measurements confirm that the elasto-bubble preserves its initial volume within experimental uncertainty over the duration of the experiments.

\begin{figure}
\includegraphics[width=0.9\linewidth]{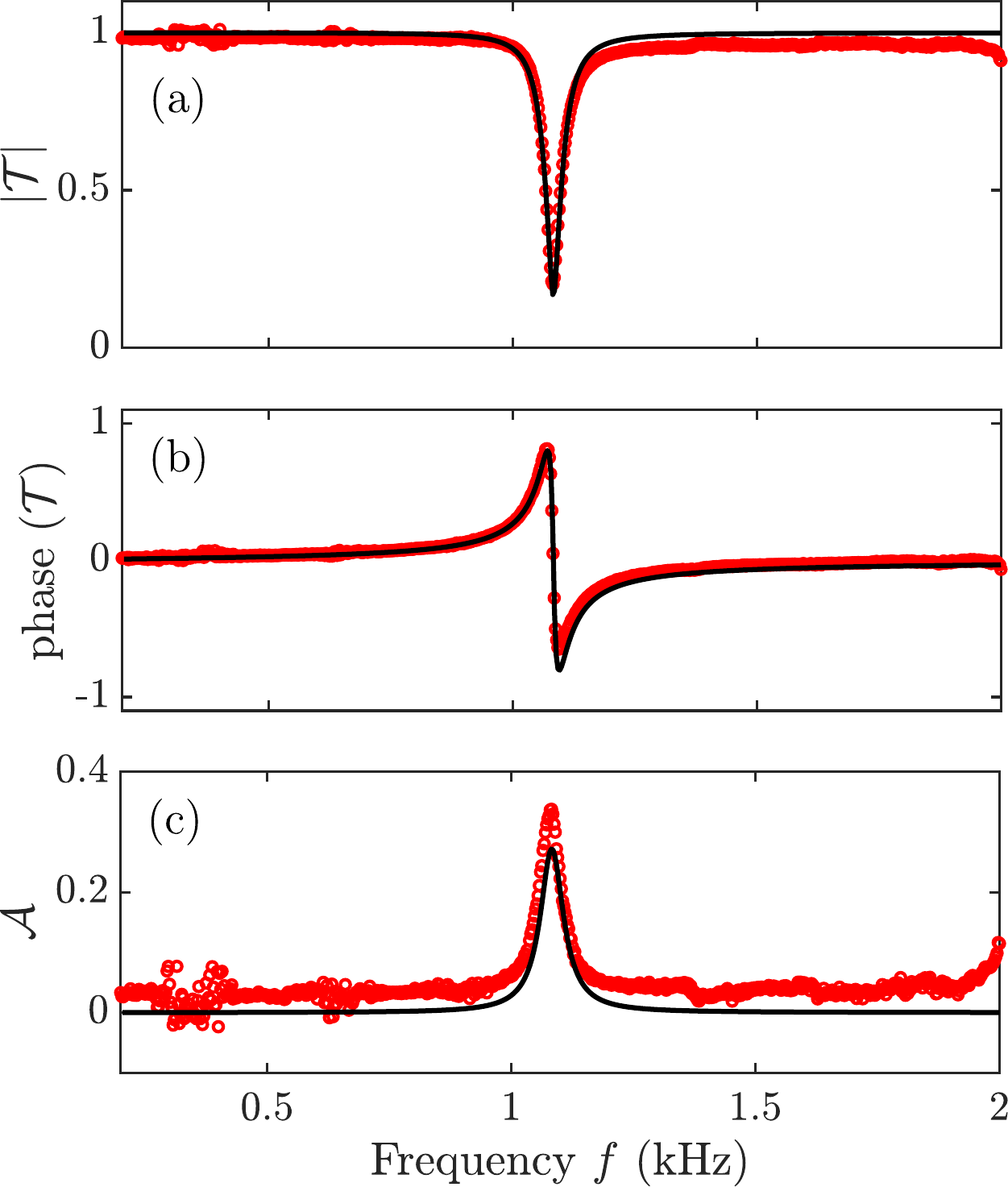}
\caption{\label{fig:TandA} Example of wave transmission amplitude (a), phase (b), and absorption (c) for the elasto-bubble shown in Fig.~\ref{fig:schema}. Red circles: measurements; solid black curves: model prediction including the rheological law~\cite{delory2022soft}. No adjustable parameter is used.}
\end{figure}

Figure~\ref{fig:TandA}(a,b) shows the transmission amplitude and phase, respectively $|\mathcal{T}|$ and $\mathrm{phase}(\mathcal{T})$, measured for elasto-bubble No.~2 in the impedance tube. A sharp drop in transmission is observed at $f_\mathrm{min}=1.1\,$kHz, together with a pronounced phase feature, indicating a strong resonant interaction between the acoustic wave and the elasto-bubble. Figure~\ref{fig:TandA}(c) shows the corresponding absorption, defined as $\mathcal{A}=1-|\mathcal{R}|^2-|\mathcal{T}|^2$, which reaches a maximum of $0.34$.

The coincidence between the transmission anomaly and the resonance frequency predicted by the model ($f_0=1.06\,$kHz for sample No.~2; Table~\ref{tab:table}) already suggests that the elasto-bubble behaves as expected. A more quantitative comparison is obtained by calculating the transmission and reflection coefficients predicted by the theory. For a monopole in a square duct, image monopoles can be used to satisfy the zero-displacement boundary conditions at the walls. This is equivalent to considering an infinite plane of monopoles arranged on a square lattice and leads to~\cite{leroy2009transmission,skvortsov2019sound}
\begin{eqnarray}
\mathcal{R}&=& \frac{\text{i} K f_\text{scat}}{1-\text{i} Kf_\text{scat} B} \\
\mathcal{T} &=& 1+\mathcal{R} \label{eq:T1}
 \end{eqnarray} 
where $K=2/(k_a R_t^2)$ and $B=1+\mathrm{i} k_aR_t -(k_aR_t)^2/2 $. Here we assume that the circular tube gives a response close to that of a square duct with the same cross-sectional area. Injecting Eq.~(\ref{eq0}) into Eq.~(\ref{eq:T1}) yields
\begin{eqnarray}
\mathcal{T} &=& 1+\frac{\mathrm{i} K R^\star}{\left(\frac{\omega_0}{\omega}\right)^2 - 1 + \frac{2R^\star}{R_t} - \mathrm{i}(\delta + KR^\star)}. \label{eq:T2}
\end{eqnarray}

\begin{figure}[h]
\includegraphics[width=0.8\linewidth]{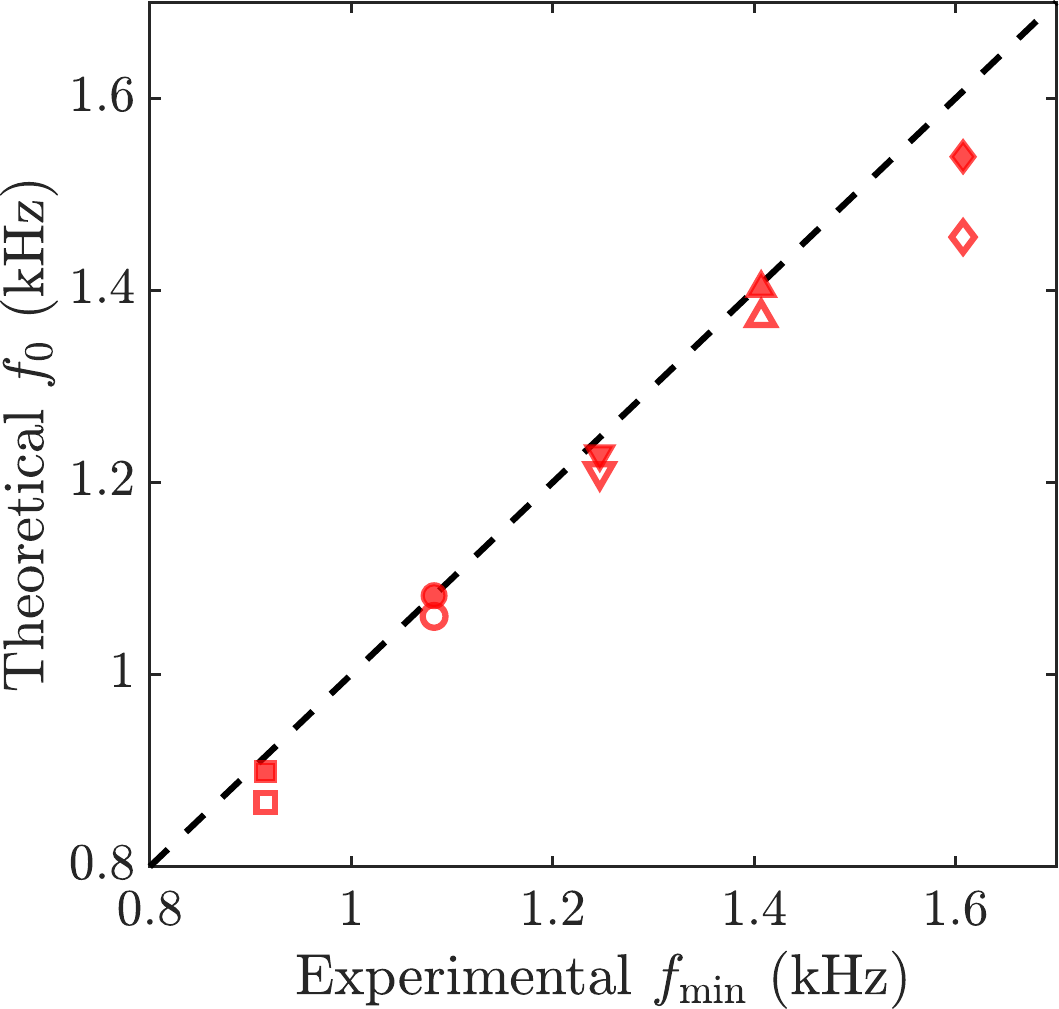}
\caption{\label{fig:f0fp} Theoretical versus experimental frequency $f_\mathrm{min}$ for the minimum of transmission. Hollow symbols correspond to the rheology-free theoretical resonance frequency given by Eq.~(\ref{eq2b}). Solid symbols show the minimum predicted by Eq.~(\ref{eq:T2}), which includes rheology and coupling effects. Black dashed line: identity line.}
\end{figure}

This equation predicts a minimum of transmission for $\omega=\omega_0(1-2R^{\star}/R_t)^{-1/2}$, i.e., slightly above the resonance frequency, due to coupling between the elasto-bubble and its images. The corresponding frequency shift is expected to remain small because the effective acoustic radius of the elasto-bubble is itself small. For elasto-bubble No.~2, $R^\star\simeq 1.1\,$mm, leading to $(1-2R^{\star}/R_t)^{-1/2}\simeq 1.02$. The shell shear modulus can also introduce deviations from the simple estimate of Eq.~(\ref{eq2b}). For Ecoflex 00-30, rheological measurements reported in the literature~\cite{delory2022soft} give $\mu=\mu_0(1-(\mathrm{i}\omega \tau)^n)$, with $\mu_0=26\,$kPa, $\tau=0.26\,$ms, and $n=0.33$. With this rheological law, the shear modulus at $1\,$kHz is of order $52\,$kPa, leading to $12\mu^\prime h/R\simeq 13\,$kPa for elasto-bubble No.~2, which remains small compared with $3\kappa P_0\simeq 420\,$kPa. The approximate resonance frequency of Eq.~(\ref{eq2b}) is therefore expected to provide a good prediction of the frequency at which the transmission reaches a minimum. Figure~\ref{fig:f0fp} shows that this expectation is reasonably verified for all of the elasto-bubbles studied here.

\begin{figure}[h]
\includegraphics[width=0.8\linewidth]{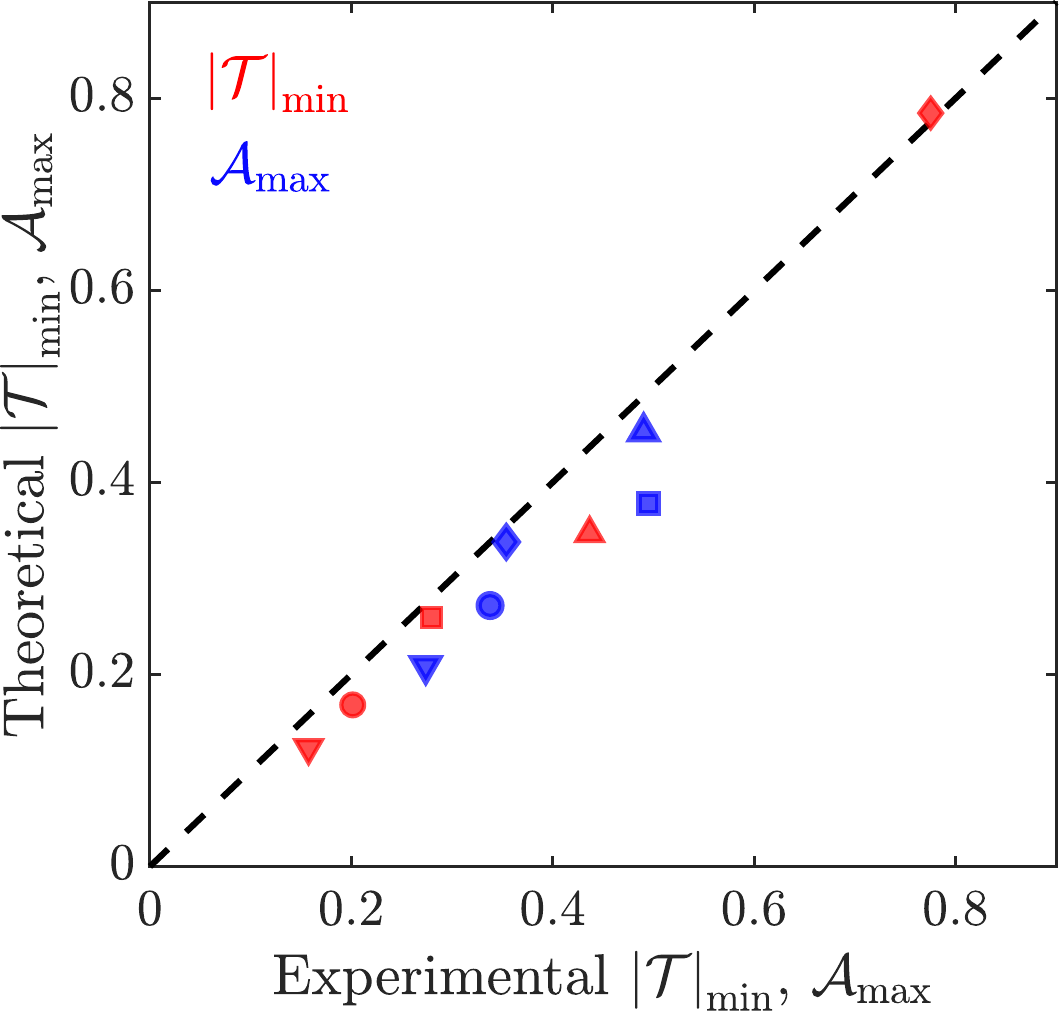}
\caption{\label{fig:TminTmodmin} Red symbols: predicted minimum transmission as a function of the measured minimum transmission $|\mathcal{T}|_{\mathrm{min}}$. Blue symbols: predicted maximum absorption as a function of the measured peak absorption $\mathcal{A}_{\mathrm{max}}$. Black dashed line: identity line.}
\end{figure}

Whereas the position of the transmission minimum depends only weakly on the rheology of the elastomer, the line shape is more sensitive to dissipation. Indeed, according to Eq.~(\ref{eq:T2}), the minimum transmission is given by $\delta/(\delta + KR^\star)$, which directly depends on the viscous losses of the shell. As shown in Fig.~\ref{fig:TandA}, the complex transmission is very well reproduced when using the literature rheological law for the elastomer.

For the absorption (Fig.~\ref{fig:TandA}(c)), the model slightly underestimates the experimental values by an approximately constant offset of about $0.03$. This discrepancy likely arises from additional losses not related to the elasto-bubble itself, since a comparable level of absorption is measured in the empty tube. A comparison between the experimental and theoretical minima of transmission and maxima of absorption is provided in Fig.~\ref{fig:TminTmodmin} for the five elasto-bubbles listed in Table~\ref{tab:table}. The overall agreement is reasonable, showing that the model efficiently captures the acoustic response of the elasto-bubbles across the full set of samples.

In summary, we have demonstrated that tailored air-filled elasto-bubbles act as airborne analogues of Minnaert bubble resonators. These objects exhibit strongly subwavelength monopolar resonances, with wavelengths of order $\lambda\sim 15R$ at resonance, and their acoustic response is quantitatively described within the framework of layered-bubble theory. Because their resonance frequency can be tuned readily through independent control of radius and shell thickness, they offer a practical route toward airborne acoustic metamaterials based on compact resonant inclusions. In the frequency range explored here, this tunability allows resonances to be shifted from about $800\,$Hz to $1455\,$Hz, while maintaining strong transmission reduction and significant absorption. These results identify elasto-bubbles as promising building blocks for acoustic filters, resonant absorbers, and metasurfaces, and more broadly extend the physics of bubble acoustics from liquid-borne to airborne environments.

\begin{acknowledgments}
\end{acknowledgments}

\section*{Data Availability Statement}

All data supporting the findings of this study are available within the manuscript. Additional raw datasets and analysis scripts used to extract wave transmission, phase, and absorption are available from the corresponding author upon reasonable request.

\bibliography{biblio}

\end{document}